\documentclass[prb,12pt]{revtex4}

\usepackage{amsmath}    
\usepackage{graphicx}   
\usepackage{verbatim}   
\usepackage{color}      
\usepackage{subfigure}  
\usepackage{hyperref}   

\begin{document}

\title{Discriminating different classes of biological networks by analyzing the graphs spectra distribution}
\author{Daniel Yasumasa Takahashi$^1$, Jo\~ao Ricardo Sato$^2$, Carlos Eduardo Ferreira$^3$, Andr\'e Fujita$^3*$}
\affiliation{
$^{1}$Department of Psychology and Neuroscience Institute, Green Hall, Princeton University, NJ 08540 Princeton, USA.\\
$^{2}$Center of Mathematics, Computation, and Cognition, Federal University of ABC, Rua Santa Ad\'elia, 166, Santo Andr\'e, 09210-170, Brazil.\\
$^{3}$Department of Computer Science, Institute of Mathematics and Statistics, University of S\~ao Paulo, Rua do Mat\~ao, 1010 - Cidade Universit\'aria, S\~ao Paulo - SP, 05508-090, Brazil.\\
$^{*}$To whom correspondence should be addressed.
}


\begin{abstract}
The brain's structural and functional systems, protein-protein interaction, and gene networks are examples of biological systems that share some features of complex networks, such as highly connected nodes, modularity, and small-world topology. Recent studies indicate that some pathologies present topological network alterations relative to norms seen in the general population. Therefore, methods to discriminate the processes that generate the different classes of networks (\textit{e.g.}, normal and disease) might be crucial for the diagnosis, prognosis, and treatment of the disease. It is known that several topological properties of a network (graph) can be described by the distribution of the spectrum of its adjacency matrix. Moreover, large networks generated by the same random process have the same spectrum distribution, allowing us to use it as a ``fingerprint". Based on this relationship, we introduce and propose the entropy of a graph spectrum to measure the ``uncertainty" of a random graph and the Kullback-Leibler and Jensen-Shannon divergences between graph spectra to compare networks. We also introduce general methods for model selection and network model parameter estimation, as well as a statistical procedure to test the nullity of divergence between two classes of complex networks. Finally, we demonstrate the usefulness of the proposed methods by applying them on (1) protein-protein interaction networks of different species and (2) on networks derived from children diagnosed with Attention Deficit Hyperactivity Disorder (ADHD) and typically developing children. We conclude that scale-free networks best describe all the protein-protein interactions. Also, we show that our proposed measures succeeded in the identification of topological changes in the network while other commonly used measures (number of edges, clustering coefficient, average path length) failed. 

\end{abstract}

\maketitle

\section{Author Summary}
There is increasing evidence that there exist tight relationships between neuronal or genetic diseases and topological changes in brain connectivity or gene regulatory networks, respectively. However, the comparison between healthy versus disease networks cannot be carried out directly by verifying for the presence or absence of each interaction, because there are topological differences within healthy people and also within patients. Even people belonging to the same group present different neuronal or genetic topological features in their networks that make them unique. Therefore, it becomes crucial to develop methods that are able to compare not just the topological features of the network, but that can verify whether two networks are generated by the same process or not. To this end, we developed statistical methods that succeeded in the identification of differences between typically developing children and those diagnosed with Attention Deficit Hyperactivity Disorder. The same set of methods was used to decide whether protein-protein interaction networks of different species are better described by Erd\"os-Renyi, scale-free, or small-world networks.

\section{Introduction}
In the last decades, attempts to understand the mechanisms that determine the topology of complex real world networks using random graphs (graphs that are generated by some random process) has gained much attention \citep{Bollobas04}. Some examples of complex networks are the World Wide Web \citep{Huberman99}, human social networks \citep{Wasserman94}, protein-protein interaction networks \citep{Maslov02}, metabolic networks \citep{Hartwell99}, and brain connectivity networks \citep{Eguiluz05}. On studying these complex networks, some questions naturally arise. For example, how complex is a given random graph? How different are two random graphs? Given a realization of a random graph, how can one infer which random graph processes generated it? Attempts to answer some of these questions have been made on purely theoretical grounds \citep{Mieghem11}, but interestingly, to the best of our knowledge, no simple and robust procedure exists to answer these questions using empirical data sets. Our aim in this work is to introduce such procedures.

Interactions are essential to understand complex systems where, to determine the behavior of the system, it is important to understand the way each component of the system interacts with others. For most classes of complex systems, interactions are neither invariant in time nor across systems from the same class. For example, neural networks in the cortex of the same individual can change in time, and synaptic organization is different among individuals. Therefore, a search for an exact common network structure seems to be unfruitful. What seems to be invariant are some statistical features that can be reproduced in classes of random graphs; therefore, the corresponding ensemble of random graphs can be used as a plausible model for an ensemble of cortical networks.


Two random graph models that are widely used to model natural phenomena are the scale-free \citep{Barabasi99} and the small-world networks \citep{Watts98}. The main characteristics of these random graphs are the non-trivial topological features that differ from the Erd\"os-R\'enyi random graphs \citep{Erdos59}, i.e., complex networks present heavy tail in the degree distribution, high clustering coefficient, community, and hierarchical structures and short path lengths. Usually, the scale-free network is characterized by its power-law degree distribution while the small-world network presents short path length and high clustering.  However, although these characteristics are essential features of these random graphs, they are not sufficient to unambiguously identify a graph as belonging to a particular class. For example, small-world networks are highly clustered like regular lattices and have small characteristic path lengths like Erd\"os-R\'enyi random graphs.

In this work we propose that the random graph spectrum, i.e., the ensemble average of the eigenvalues of the adjacency matrix, is a better and more general characterization of complex networks in comparison with other commonly used measures: number of edges, clustering coefficient, and average path length. For instance, it is known that several topological properties of a random graph, such as the number of walks, diameter, and cliques can be described by the spectrum of its adjacency matrix \citep{Mieghem11}. Based on this relationship between the topological properties of the random graph and its spectrum, we introduce the definition of entropy of a random graph spectrum and the Kullback-Leibler divergence between two random graph spectra.
By simulation experiments, we observe that the entropy of random graph spectrum is related to the intuitive idea of amount of uncertainty of a random graph and that the Kullback-Leibler divergence between random graph spectra can discriminate two random graphs that were generated by different random process. 

Statistical approaches such as model selection, parameter estimation, and hypothesis testing to discriminate two classes of random graphs are also presented. We illustrate practical use of the model selection approach in protein-protein interaction networks of eight different species. By analyzing the random graph spectrum instead of the degree distribution, we classified all the eight protein-protein interaction networks as scale-free graphs. Finally, the power of Kullback-Leibler based statistical test is illustrated by an application in networks derived from children with Attention Deficit Hyperactivity Disorder and with typical development.  We succeeded in the identification of topological changes between children with typical development and ADHD patients, while standard measures such as number of edges, clustering coefficient and average path length failed.

\vspace{0.7cm}
\noindent {\bf Definition of graphs and graph spectrum}\\
A {\it graph} is a pair of sets $G = (P,E)$, where $P$ is a set of $n$ nodes and $E$ is a set of $m$ edges that connect two nodes (elements of $P$). A random graph $g$ is a family of graphs, where each member of the family is generated by some probability law. Among several classes of random graphs, there are three that have known importance due to their capability to model real world events, namely, Erd\"os-R\'enyi random (Figure 1a) \citep{Erdos59}, scale-free (Figure 1b) \citep{Barabasi99}, and small-world graphs (Figure 1c) \citep{Watts98}.

\begin{figure}[!tpb]
\includegraphics[angle=0, width=1\textwidth]{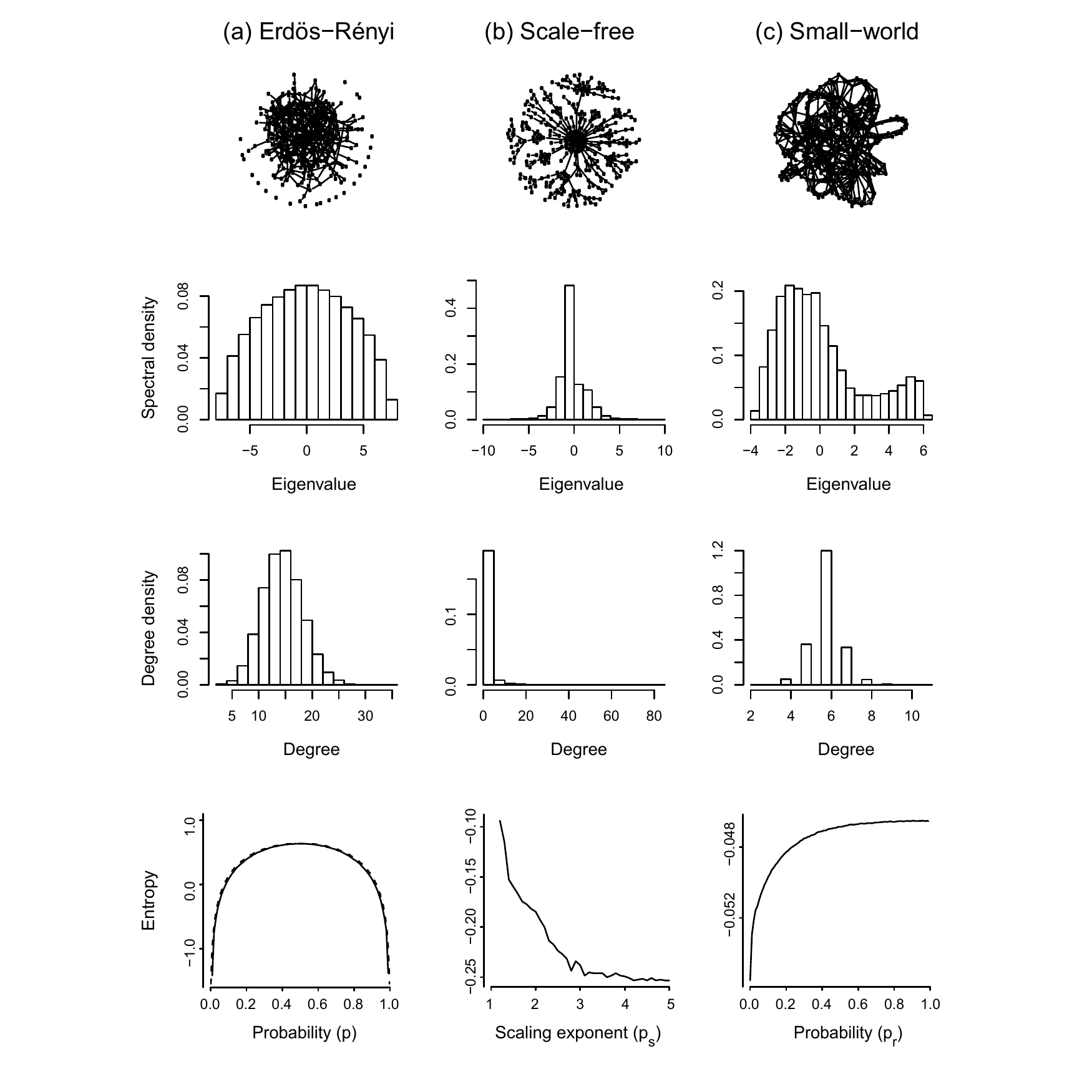}
\caption{Illustrative figure of the three different complex network models (a) Erd\"os-R\'enyi; (b) Scale-free; and (c) Small-world and their respective spectra, degree distributions, and entropies, in this order from top to bottom. The estimated entropies are computed for the respective graph type for the respective parameters (probability $p$ for the  Erd\"os-R\'enyi, scaling exponent $p_s$ for the scale-free, and probability $p_r$ for the small-world random graphs). In (a) the entropy values estimated from the simulation data is depicted by a solid line and the theoretical value of the entropy computed using equation \ref{eq:theoent} is indicated by a dashed line. 
}\label{fig1}
\end{figure}

Erd\"os-R\'enyi random graphs are the simplest ones in terms of construction. Erd\"os and R\'enyi define a random graph as $n$ labeled nodes in which each pair of nodes $(i,j)$ is connected by an edge with a given probability $p$.

Scale-free networks, proposed by Barab\'asi and Albert (1999), have a power-law degree distribution due to node preferential attachment. Barab\'asi and Albert (1999) proposed the following construction of a scale-free network: start with a small number of $(n_{0})$ nodes and at every time-step, add a new node with $m_{1} (\le n_{0}) $ edges that link the new node to $m_{1}$ different nodes already present in the system. When choosing the nodes to which the new node connects, assume that the probability that a new node will be connected to node $i$ is proportional to the degree of node $i$ and the scaling exponent $p_{s}$ which indicates the order of the proportionality ($p_{s}=1$ linear, $p_{s}=2$ quadratic and son on).

Small-world graphs, proposed by Watts and Strogatz (1998) are one-parameter models  that interpolate between a regular lattice and an Erd\"os-R\'enyi random graph \citep{Newman99}. First, a ring lattice with $n$ nodes is constructed, in which every node is connected to its first $K$ neighbors ($K/2$ on either side). Then, we choose a vertex and the edge that connects it to its nearest neighbor in a clockwise sense. With probability $p_s$ we reconnect this edge to a vertex chosen uniformly at random over the entire ring. This process is repeated by moving clockwise around the ring, considering each vertex in turn until one lap is completed. Next, the edges that connect vertices to their second-nearest neighbors clockwise are considered. As the previous step, each edge is randomly rewired with probability $p_s$, and continue this process, circulating around the ring and proceeding outward to more distant neighbors after each lap, until each edge in the original lattice has been considered once \citep{Watts98}.

Any undirected graph $G$ with $n$ nodes can be represented by its adjacency matrix $A(G)$ with $n\times n$ elements $A_{ij}$, whose value is $A_{ij}=A_{ji}=1$ if nodes $i$ and $j$ are connected, and 0 otherwise. The {\it spectrum} of graph $G$ is the set of eigenvalues of its adjacency matrix $A(G)$. A graph with $n$ nodes has $n$ real eigenvalues $\lambda_{1}\ge\lambda_{2}\ge \ldots \ge \lambda_{n}$. Now, given a random graph $g$, the eigenvalues  are random vectors for which we can take the expectation with respect to the probability law of the random graph. We define the {\it spectral density} distribution of a random graph $g$  as

\begin{equation}
\rho_{g}(\lambda)=\lim_{n\rightarrow\infty}\left<\frac{1}{n}\sum_{j=1}^{n}\delta(\lambda-\lambda_{j}/\sqrt{n})\right>,
\end{equation}
where $\delta$ is the Dirac delta function and the brackets ``$\langle \rangle$'' indicate the expectation with respect to the probability law of the random graph. In what follows, we use the shorthand name spectrum of $g$ to indicate $\rho_{g}$. The interest in spectral properties is related to the fact that the spectral density can be directly related to the graph's topological features \citep{Albert02}. 

In application, a closed form for the spectral density is rarely available, so we have to rely on some statistical estimators $\hat{\rho}_{g}$. In order to estimate the spectral densities, first the eigenvalues are computed, and then Gaussian kernel regression using the Nadaraya-Watson estimator \citep{Nadaraya64} is applied for the regularization of the estimator. Finally, the density is normalized to obtain the integral below the curve equal to one. The bandwidth of the kernel can be chosen by (max(eigenvalues) - min(eigenvalues))/number of bins \citep{Sain96}, where the number of bins can be selected by using any objective criterion. In this work, we used the Sturges' criterion \citep{Sturges26}. 

It is worth mentioning that the study of spectral density distribution of complex networks is still an active area of research \citep{Mieghem11, Dorogovtsev03}, but the aim has been in general to obtain the exact or approximate  properties of spectrum distribution for a given model. In this article, we are instead concerned with their statistical properties and  applications to crucial biological systems.

\section{Results}
First we will present the definitions of entropy and divergence for graphs spectra, along with statistical methods for estimation and significance testing. Then, the performance of each method is evaluated by simulations and finally applied to actual data for illustration.

\subsection*{Entropy of graph spectrum}
Let $\rho_{g}$ be the spectrum of a random graph $g$. We define the {\it spectral entropy} $H(\rho_{g})$ as

\begin{equation} \label{eq:ent}
H(\rho_{g})=-\int_{-\infty}^{+\infty}\rho_{g}(\lambda)\log\rho_{g}(\lambda)d\lambda,
\end{equation}
where, as usual, we assume $0\log 0 = 0$. Observe that the entropy defined above is also known as differential entropy \citep{Cover2006} and can assume negative values, in contrast to the entropy defined for discrete distributions.
	
	Since the spectral density of an adjacency matrix of a random graph has a tight relationship with the random graph structure and can be considered a fingerprint of the random graph \citep{Mieghem11}, we propose that the corresponding spectral entropy also describes important characteristics of the random graph. More specifically, we propose that the spectral entropy measures a form of ``uncertainty" associated to the random graph.
	To gain some intuition, we can compute the approximate spectral entropy for the Erd\"os-R\'enyi random graph $g$ with parameter $p$ as follows. For large $n$, we have
\begin{equation}
\rho_{g}(\lambda)\sim\frac{\sqrt{4p(1-p)-\lambda^{2}}}{2\pi p(1-p)}
\end{equation}
for $0<|\lambda|<2\sqrt{(p(1-p))}$ and 0 otherwise \citep{Wigner55, Wigner58}. Using the above approximation, we have that
\begin{equation} \label{eq:theoent}
H(\rho_{g} )\sim\frac{1}{2}\ln(4\pi^{2}p(1-p))-\frac{1}{2}.
\end{equation}

	This formula shows that the maximum spectral entropy for the Erd\"os-R\'enyi graph is achieved for $p=0.5$, which is in accordance to the intuition that this is the model with the largest uncertainty. To confirm our point, the Erd\"os-R\'enyi random graph spectral entropy was calculated for many different values of probability $p$ (bottom panel Figure 1a, dashed line). For the Erd\"os-R\'enyi graphs, not surprisingly, the entropy achieved its maximum value on $p=0.5$, and the minimum values on $p=0$ and $p=1$, which is the situation where there is only one possible graph, i.e., the empty and complete graphs, respectively (Figure 1a). Furthermore, it is important to point out that the entropy function is symmetric due to the symmetry of the spectrum function, i.e., the spectral density of the Erd\"os-R\'enyi graph generated with parameter $p$ is equal to the spectral density of the Erd\"os-R\'enyi graph generated with parameter $1-p$.

For the scale-free and small-world networks, an exact formula for the spectral entropy is not known, therefore, we estimated the entropy for different parameters of the models. A straightforward way to obtain an estimator $\hat{H}(\rho_g)$  for the spectral entropy is to first obtain an estimator $\hat{\rho}(\lambda)$ of $\rho(\lambda)$ and plug in to the equation (2). This is the procedure adopted in this work. To verify the accuracy of our estimator we compared the average estimated entropy values for 100 Erd\"os-R\'enyi random graphs with 500 nodes (bottom panel Figure 1a, solid line) and the theoretical value in equation \ref{eq:theoent} (bottom panel Figure 1a, dashed line). A visual inspection shows that the estimator is very accurate. The average bias for this example was $-0.015$, i.e., a small negative bias. 

For the scale-free graphs we observe (Figure 1b) that the estimated entropy is higher in low scaling exponents ($p_{s}$) because it becomes similar to an Erd\"os-R\'enyi random graph, whereas when the scaling exponent goes to infinity it becomes closer to a complete bipartite graph resulting in a lower entropy. Finally, for small-world graphs (Figure 1c), the entropy is higher when the randomness of the graph (probability $p_{r}$) increases. Notice that when $p_{r}=1$, the small-world graph becomes an Erd\"os-R\'enyi graph, whereas when $p_r = 0$ the graph is a ring \citep{Watts98}, therefore presenting lower entropy. For both scale-free and small-world graphs, the number of nodes and edges were set to 500 and 600, respectively, and for each scaling exponent ($p_{s}$) or probability ($p_{r}$), an average entropy of 100 graphs were calculated.

\subsection*{Kullback-Leibler divergence between graphs}
Once the spectral entropy is defined, one may introduce a measure of similarity between two spectral densities, which is also a measure of similarity between two random graphs. It is clear that if two spectral densities are different, then the respective graphs should be different, although the converse is not always true (i.e., there are non-isomorphic graphs which are isospectral).

We define the {\it Kullback-Leibler divergence} (for sake of brevity we call it KL divergence) between two spectral densities $\rho_{g_{1} }$ and $\rho_{g_{2} }$ as

\begin{equation}
KL(\rho_{g_{1} } |\rho_{g_{2}} )=\int_{-\infty}^{+\infty}\rho_{g_{1}}(\lambda)\log\frac{\rho_{g_{1}}(\lambda)}{\rho_{g_{2}}(\lambda)}d\lambda,
\end{equation}
if the support of $\rho_{g_{2} }$ contains the support of $\rho_{g_{1} }$. Otherwise, $KL(\rho_{g_{1} } |\rho_{g_{2}} ) = +\infty$. As usual, we assume $0 \log \frac{0}{0} = 0$.

For the above equation, $\rho_{g_{2}}$ is called the {\it reference measure}. This divergence is asymmetric and non-negative. It is also zero if and only if $\rho_{g_{1}}$ and $\rho_{g_{2}}$ are equal.

	The KL divergence can be interpreted as a measure of discrepancy between two random graphs, thus can be used to build an estimator for the parameter of a model given an observation. Specifically, let $g$ be a random graph with spectral density $\rho_{g}$. Also let $\{\rho_{\theta}\}$ be a parametric family of spectral distributions indexed by a real vector $\theta$. Assume that there exists a value of the parameter $\theta$, which we denote $\theta^{*}$ that minimizes $KL(\rho_{g} |\rho_{\theta})$. An estimator $\hat{\theta}$ of $\theta^{*}$ is given by 

\begin{equation} \label{eq:est}
\hat{\theta}=\arg\min_{\theta}KL(\hat{\rho}_{g}|\rho_{\theta}).
\end{equation}
The idea is that among all possible choices of models in a parametric class of random graphs $\rho_{\theta}$, we choose the one for which the corresponding spectral density minimizes the divergence with the non-parametrically estimated spectral density. This is in the same spirit as nonparametric likelihood estimators of which the Whittle estimator is an example \citep{Whittle53}.

To show the performance of our estimator, different complex network models (Erd\"os-R\'enyi, scale-free, and small-world) with sizes equal to 50, 100, 200, and 300 nodes were simulated. The parameters to be estimated for each random graph model are: the probability $p$ of connecting two nodes for Erd\"os-R\'enyi graphs, the scaling exponent of the preferential attachment $p_{s}$ for scale-free graphs, and the rewiring probability $p_{r}$ for small-world graphs. The estimated parameters were averaged values calculated for 50 repetitions, and the results are shown in Table \ref{table1}. Brackets indicate one standard deviation. From the results in Table \ref{table1}, we conclude that the estimator is reasonable and it can recover the correct parameter with relatively small bias and variance, i.e., one or two order of magnitudes smaller than the value of the estimated parameter. We observe from Table \ref{table1} and further simulations not shown here that the direction of the bias depend on the specific parameter of the model and size of the graph, and therefore no systematic bias direction seems to exist. The performance of the estimator is further discussed in Section \ref{sec:discussion}

\begin{table}[ht]
\caption{Average parameters estimated by minimum distance estimator based on KL divergence for Erd\"os-R\'enyi random, scale-free, and small-world graphs.  One standard deviations are indicated between brackets. Calculations were carried out for 50 repetitions. The parameters to be estimated for each graph model are: the probability $p$ of connecting two nodes for Erd\"os-R\'enyi graphs, the power of the preferential attachment $p_{s}$ for scale-free graphs, and the rewiring probability $p_{r}$ for small-world graphs.}
\centering
\begin{tabular}{c c c c}
\hline\hline
& Random ($p$) & Scale-free ($p_{s}$) & Small-world ($p_{r}$)\\
Number of nodes / true parameters & 0.50 & 1.50 & 0.30\\
50 & 0.51 (0.04) & 1.53 (0.06) & 0.33 (0.05)\\
100 & 0.50 (0.03) & 1.53 (0.05) & 0.33 (0.03)\\
200 & 0.50 (0.03) & 1.56 (0.03) & 0.34 (0.03)\\
300 & 0.50 (0.03) & 1.55 (0.05) & 0.34 (0.03)\\
500 & 0.50 (0.02) & 1.54 (0.04) & 0.33 (0.03)\\ 
\hline
\hline
\end{tabular}
\label{table1}
\end{table}

Another use of the KL is to build a model selection criterion to select good models among a set of candidate random graphs. More specifically, given a graph, it is important to decide if the graph can be better predicted by an Erd\"os-R\'enyi, scale-free, or small-world networks. The KL divergence between the given graph spectrum and the spectrum of different classes of graphs can be interpreted as the quality of fitting the graph to the model. 

Given a graph $g$ and its spectrum $\rho_g$, several candidate graph models may be ranked according to their KL divergence values and the models with smaller KL divergence values should be considered as good candidates to explain the data. Thus, KL divergence provides an objective comparison among models, i.e., a tool for model selection. Specifically, let $\hat{\rho}_g$ be the empirical spectral distribution and $\{\rho_{\theta_1}\}, \ldots, \{\rho_{\theta_m}\}$ be $m$ different parametric families of spectral distributions. Let $\hat{\theta}_i$ for $i = 1, \ldots, m$ be the estimators given in  equation \ref{eq:est}. We denote by $\#(\theta_i)$ the dimension of $\theta_i$. The best candidate model $\hat{\theta}_j$ is chosen by

\begin{equation} \label{eq:modelselec}
j = \arg\min_{i} 2KL(\hat{\rho}\vert\rho_{\hat{\theta}_i}) + 2\#(\hat{\theta}_i)
\end{equation}

The motivation for this criterion is the AIC (Akaike Information Criterion) \citep{Akaike74} model selection criterion. Informally, the model that minimizes equation \ref{eq:modelselec} is the one that has the most similar spectral distribution when compared to the spectral distribution of the data. The penalization term $2\#(\hat{\theta}_i)$ is added to avoid overfitting. The three random graph models analyzed here have the same number of parameters; therefore, the penalization term is not required.

Simulations were carried out in order to verify the accuracy of the proposed model selection approach. Ten thousand graphs of each class were generated and classified as Erd\"os-R\'enyi, scale-free, or small-world by the model selection approach. The graph size varied from 10 to 120 nodes. Figure 2 illustrates the performance of the model selection method.
For all graph class (Erd\"os-R\'enyi (Figure 2a), scale-free (Figure 2b) or small-world (Figure 2c)), when the number of nodes increases, the correct proportion of hits also increases, demonstrating that the method is consistent and improves with the graph size.

Usually, in real applications, complex networks are composed of hundreds to thousands of nodes. In Figure 2, we observe that the accuracy is high even for graphs smaller than 100 nodes. Indeed, this implies that the proposed model selection method should be useful for applications in data set with realistic data size.

\begin{figure}[!tpb]
\includegraphics[angle=0, width=1\textwidth]{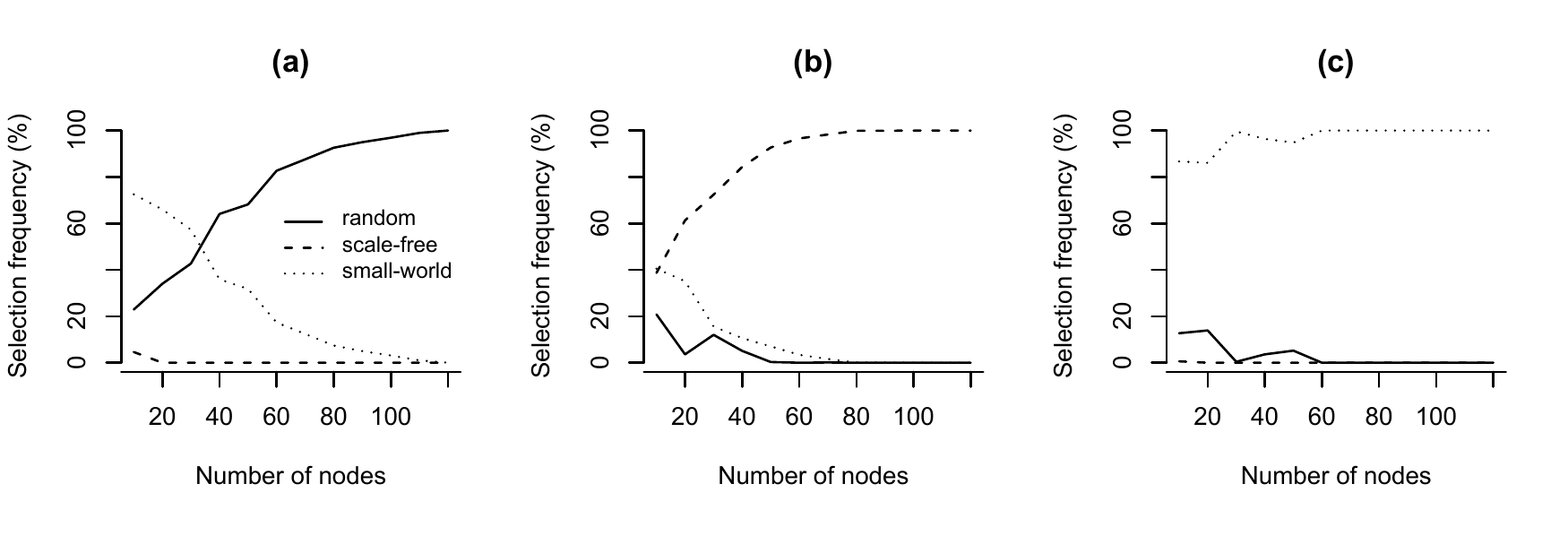}
\caption{Figure illustrating the performance of the model selection approach. Given a graph belonging to (a) Erd\"os-R\'enyi with parameter $p=0.3$, (b) scale-free with parameter $p_{s}=1$, and (c) small-world with parameter $p_{r}$=0.3, the solid, dashed, and dotted lines represent the proportion of graphs classified as Erd\"os-R\'enyi, scale-free, and small-world, respectively. Notice that the larger is the graph, the higher is the proportion of correct hits, showing that the model selection approach is consistent. For each graph size (10, 20, 30, 40, 50, 60, 70, 80, 90, 100, 110, 120 nodes), 1,000 repetitions were carried out.
}\label{fig2}
\end{figure}

Interestingly, the performance to identify small-world graphs is very high, close to 100\% even when the graph is very small (10 nodes). This is  probably due to the specific algorithm to construct such a graph. Remember that the construction of a small-world graph based on Watts-Strogatz algorithm starts with a deterministic step, i.e., a ring lattice with $n$ nodes which every node is connected to its first $K$ neighbors ($K/2$ on either side). It is likely that this first step makes this type of graph different whether compared to Erd\"os-R\'enyi or scale-free graphs that are totally non-deterministic.

\subsection*{Jensen-Shannon divergence}
Given two random graphs $g_1$ and $g_2$, now we would like to define a notion of distance between them based on entropy. In other words, we are interested in identifying graphs that are generated by the same random process instead of isomorphism in graphs (an isomorphism of graphs $g_1$ and $g_2$ is a bijection $f$ from the vertex sets of $g_1$ to the vertex sets of $g_2$ such that any two vertices $u$ and $v$ of $g_1$ are adjacent if and only if $f(u)$ and $f(v)$ are adjacent in $g_2$)

The KL divergence is suited for the purpose of parameter estimation and model selection as explained in previous section. Nevertheless, it is not symmetric, i.e., in general $KL(\rho_1 | \rho_2) \neq KL(\rho_2 | \rho_1)$. For this reason, KL divergence is not suited when it is not clear which distribution is the reference distribution. This is indeed the case for statistical test comparing two graphs spectra $\rho_1$ and $\rho_2$. We would like to avoid inconsistency in the results when considering $KL(\rho_1 | \rho_2)$ or $KL(\rho_2 | \rho_1)$.

Therefore, we introduce the {\it Jensen-Shannon divergence} (JS) between two spectral densities $\rho_{g_{1}}$ and $\rho_{g_{2}}$ defined as 
\begin{equation}
JS(\rho_{g_{1}}, \rho_{g_{2}})=\frac{1}{2}KL(\rho_{g_{1}}\vert \rho_{M}) + \frac{1}{2}KL(\rho_{g_{2}}\vert \rho_{M})
\end{equation}
where $\rho_{M}=\frac{1}{2}(\rho_{g_{1}}+\rho_{g_{2}})$.

This divergence is symmetric and non-negative. It is also zero if and only if $\rho_{g_{1}}$ and $\rho_{g_{2}}$ are equal. Moreover, the square root of the JS divergence satisfies the triangle inequality.

It is natural to ask if the JS divergence between two distributions is zero or not. Therefore, we set the statistical test for JS divergence between two sets of graphs spectra $\rho_{g_{1}}$ and $\rho_{g_{2}}$ as ($H_{0}: JS(\rho_{g_{1}}, \rho_{g_{2}})=0$ versus $H_{1}: JS(\rho_{g_{1}}, \rho_{g_{2}})>0$). Details of the respective bootstrap-based test are provided in the Materials and Methods section.

When a statistical test is proposed, at least two properties must be shown: the power of the test under the alternative hypothesis ($H_{1}$) and the control of the rate of false positives under the null hypothesis ($H_0$).

In order to check the power of the statistical test, i.e., if the method based on the spectral distribution actually discriminates between two sets of graphs characterized by slightly different parameters (details in the Materials and Methods section), ROC (Receiver Operating Characteristic) curves were constructed and compared to the test based on the degree distribution. The ROC curve is useful in evaluating the power of the test and it consists in a bidimensional plot of sensitivity (y-axis) versus 1 - specificity (x-axis), where sensitivity = number of true positives/(number of true positives+number of false negatives) and specificity = number of true negatives/(number of true negatives + number of false positives). The area below the ROC curve is a quantitative summary of the power of the test. In other words, an area closer to one (a curve above the diagonal line) denotes high power while an area close to 0.5 (a curve close to the diagonal line) is equivalent to random decisions. The top panels in Figure 3 illustrate the ROC curves with 10,000 repetitions for each class (Erd\"os-R\'enyi, scale-free, and small-world). The solid and dashed lines represent the test based on the spectral and degree distributions, respectively. Despite the small differences between the two conditions (parameters $p_{1}=0.10$ versus $p_{2}=0.11$ for Erd\"os-R\'enyi graphs; the scaling exponent $p_{s1}=1.0$ versus $p_{s2}=1.1$ for scale-free networks; and $p_{r1}=0.30$ versus $p_{r2}=0.31$ for small-world graphs) and relatively small sizes (100 nodes), our statistical test based on the spectra was able to identify the graphs that were generated by different sets of parameters with high accuracy as can be observed by the ROC curves clearly above the diagonal line. On the other hand, the statistical test based on the degree distribution was equivalent to the spectra-based test only when the evaluated networks were Erd\"os-R\'enyi graphs. When the degree-based test was applied to scale-free and small-world graphs, the discriminative power was not much better than by chance, i.e., the ROC curves were close to the diagonal. This probably occurred because the degree distribution is closely related to the number of edges while the spectrum is related to the whole structure of the graph. Notice that the parameter $p$ of the Erd\"os-R\'enyi graph is associated to the number of edges, while the parameters $p_s$ of the scale-free network and $p_r$ of the small-world network are associated to the structure of the graph.

\begin{figure}[!tpb]
\includegraphics[angle=0, width=1\textwidth]{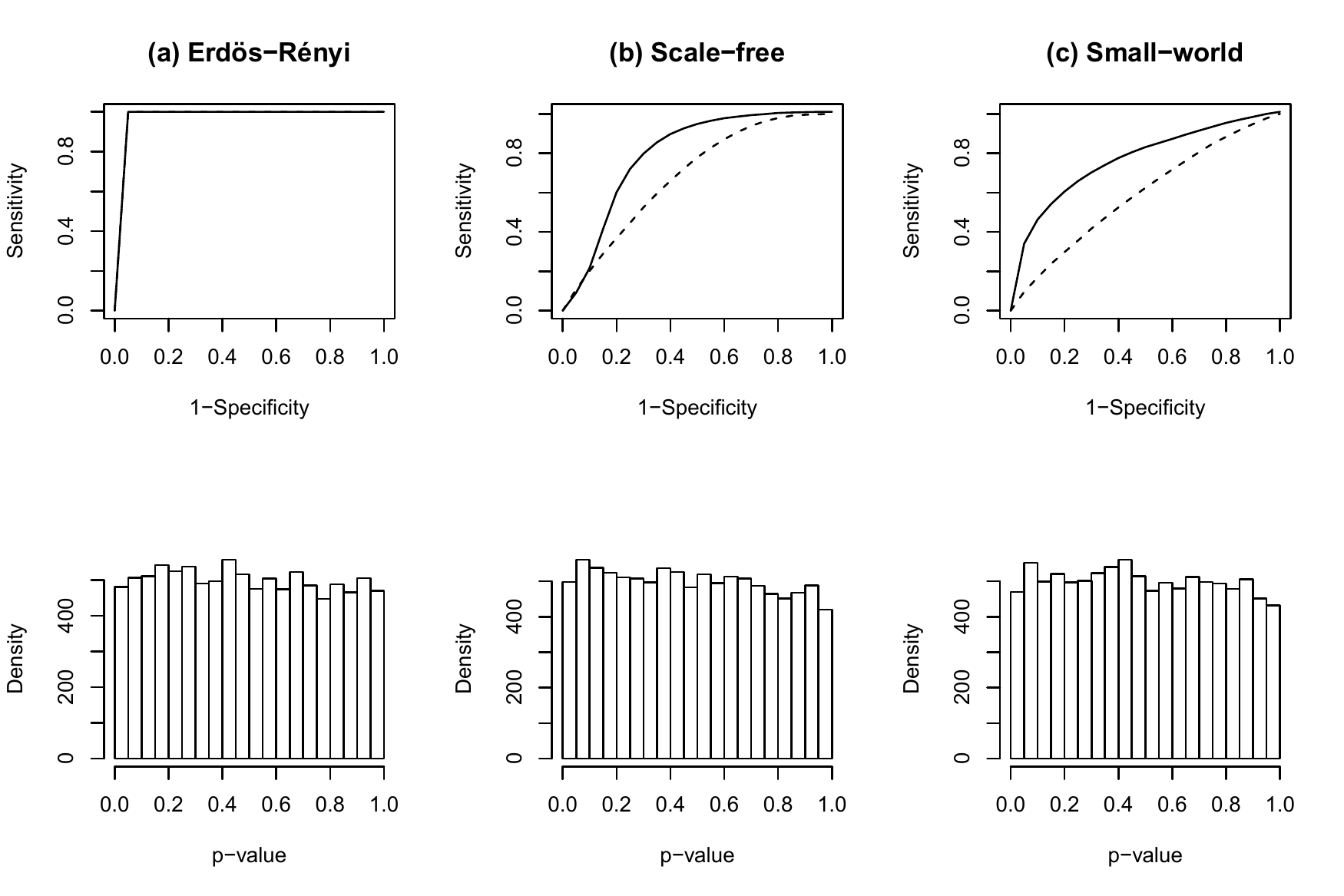}
\caption{ROC curve under the alternative hypothesis and p-value distribution under the null hypothesis for (a) Erd\"os-R\'enyi graphs; (b) scale-free graphs, and (c) small-world graphs. Both ROC curves and p-value distributions were constructed by analyzing 10,000 experiments. Solid and dashed lines represent the test based on the spectral and degree distributions, respectively.
}\label{fig3}
\end{figure}

It is also necessary to verify if the bootstrap-based test is actually controlling the rate of false positives under the null hypothesis, i.e., when both sets of graphs are generated by the same model and same set of parameters. By simulating two random graphs $g_{1}$ and $g_{2}$, each one generated by the same model and parameters (see Materials and Methods section), and testing  $H_{0}: JS(\rho_{g_{1}}, \rho_{g_{2}})=0$ versus $H_{1}: JS(\rho_{g_{1}}, \rho_{g_{2}})>0$, the p-value distribution should be a uniform distribution. The uniform distribution of p-values illustrates that the rate of false positives is actually controlled under any p-value threshold, since the uniform distribution emerges for p-values when the distribution of the null hypothesis is correctly specified by our bootstrap procedure. Notice that for a p-value threshold set to 1\%, it is expected to obtain 1\% of false positives, for a threshold of 5\%, 5\% are expected to be false positive and so on and so forth. The bottom panels in Figure 3 show the p-value distributions (x-axis represents the p-values while the y-axis is the frequency or density of the respective p-value in 10,000 repetitions under the null hypothesis), one for each class (Erd\"os-R\'enyi, scale-free, and small-world), indicating that all of them are very similar to uniform distributions on $[0,1]$ under the null hypothesis. In other words, the bootstrap test is controlling the rate of false positives, as expected.

\subsection*{Application to protein-protein interaction network}
In order to illustrate the model selection application in actual data, protein-protein interaction data were downloaded from the DIP (Database of Interacting Proteins - \href{http://dip.doe-mbi.ucla.edu/dip/}{http://dip.doe-mbi.ucla.edu/dip/}) on June 29th, 2011. The DIP database is composed of eight species namely, {\it H. pylori} (bacterium), {\it R. norvegicus} (rat), {\it M. musculus} (mouse), {\it E. coli} (bacterium), {\it C. elegans} (worm), {\it S. cerevisiae} (yeast), {\it H. sapiens} (human), {\it D. melanogaster} (fruit fly). All of them present different number of nodes, edges, average degree, diameter, clustering coefficient and average path length as can be visualized in Table \ref{tableSpecies}. The adjacency matrices of graphs were constructed for each species and the set of eigenvalues with the corresponding multiplicities were calculated. The spectral distributions of the eight species are displayed in Figure 4.

\begin{table}[ht]\scriptsize
\caption{The general characteristics of eight protein-protein interaction networks. For each network we indicate the number of nodes, the number of edges, the average degree, the diameter, the clustering coefficient and the average path length.}
\centering
\begin{tabular}{c c c c c c c}
\hline\hline
Species & Number of nodes & Number of edges & Average degree & Diameter & Clustering coefficient & Average path length\\
{\it H. pylori} & 714 & 1,393 & 3.90 & 9 & 0.016 & 4.139\\
{\it R. norvegicus} & 758 & 691 & 1.82 & 9 & 0.001 & 3.651\\
{\it M. musculus} & 1,868 & 1,895 & 2.03 & 20 & 0.006 & 6.280\\
{\it E. coli} & 2,997 & 12,348 & 8.24 & 12 & 0.115 & 3.986\\
{\it C. elegans} & 3,183 & 5,068 & 3.18 & 13 & 0.012 & 4.803\\
{\it S. cerevisiae} & 5,213 & 25,073 & 9.62 & 10 & 0.058 & 3.860\\
{\it H. sapiens} & 5,940 & 14,144 & 4.76 & 17 & 0.017 & 4.755\\
{\it D. melanogaster} & 7,931 & 23,386 & 5.90 & 12 & 0.012 & 4.468\\
\hline
\hline
\end{tabular}
\label{tableSpecies}
\end{table}

\begin{figure}[!tpb]
\includegraphics[angle=0, width=1\textwidth]{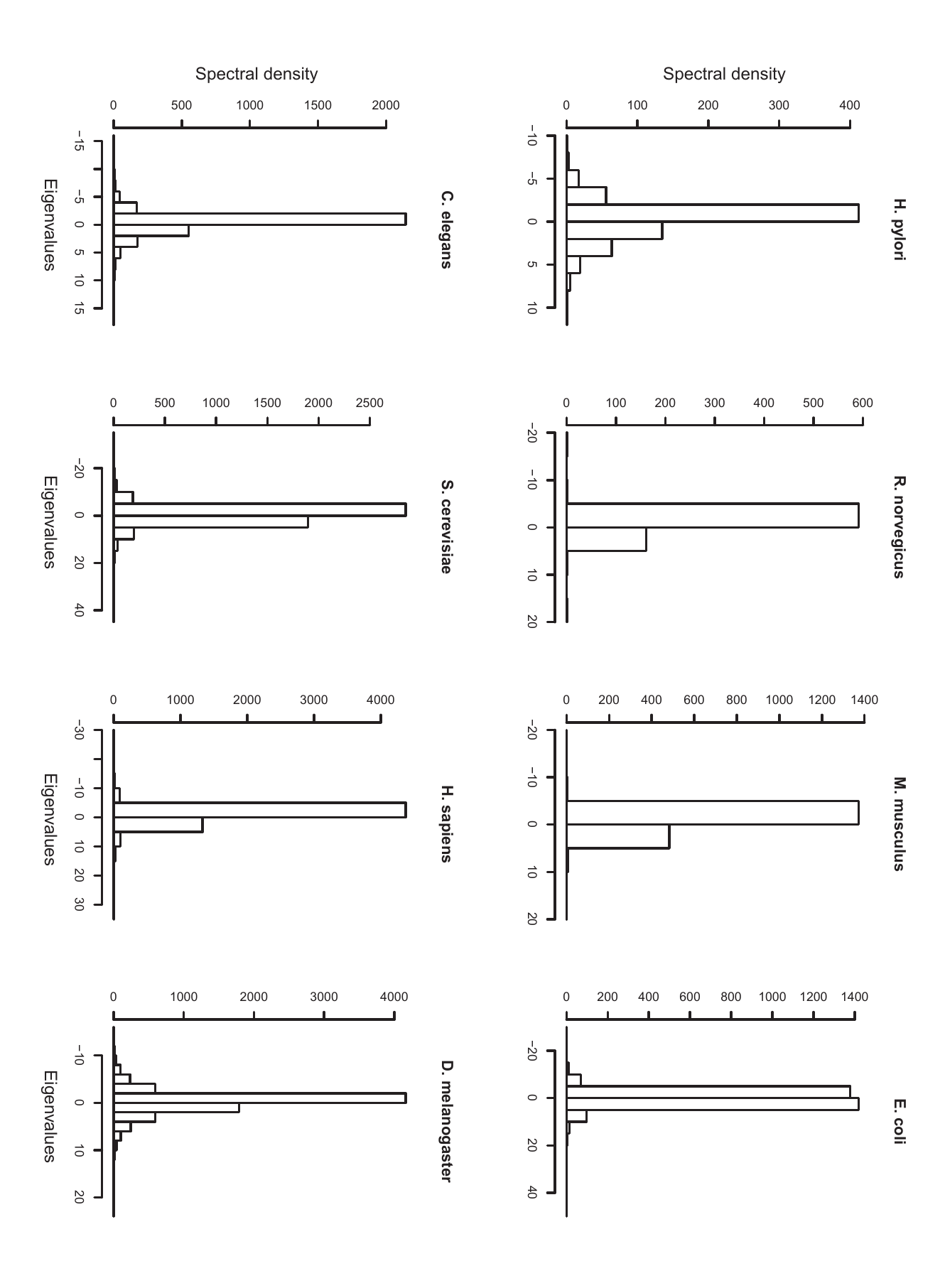}
\caption{The spectra distributions for the eight species ({\it H. pylori}, {\it R. norvegicus}, {\it M. musculus}, {\it E. coli}, {\it C. elegans}, {\it S. cerevisiae}, {\it H. sapiens}, {\it D. melanogaster}).
}\label{fig4}
\end{figure}

We evaluate how successful our algorithm based on the graph spectrum and KL divergence is by analyzing those protein-protein interaction networks that have already been classified as scale-free graphs by considering the degree distribution \citep{Jeong00}.

Remarkably, all the eight species were classified as scale-free networks by our model selection approach based on the graph spectrum analysis (instead of the degree distribution) (Table \ref{tableAIC}) demonstrating that not only the degree distribution, but also the spectrum contains information for classification.

\begin{table}[ht]\scriptsize
\caption{The estimated Kullback-Leibler divergence between the eight species and the three random graph models. In bold are the lowest KL divergence values.}
\centering
\begin{tabular}{c c c c}
\hline\hline
Species & Erd\"os-R\'enyi & Scale-free & Small-world\\
{\it H. pylori} & 15.07 & {\bf 1.46} & 11.36\\
{\it R. norvegicus} & 134.67 & {\bf 100.47} & 118.67\\
{\it M. musculus} & 14.10 & {\bf 6.93} & 24.51\\
{\it E. coli} & 21.15 & {\bf 1.91} & 17.90\\
{\it C. elegans} & 30.48 & {\bf 2.66} & 30.23\\
{\it S. cerevisiae} & 24.21 & {\bf 0.87} & 18.25\\
{\it H. sapiens} & 47.10 & {\bf 11.31} & 44.04\\
{\it D. melanogaster} & 17.40 & {\bf 0.39} & 18.06\\
\hline
\hline
\end{tabular}
\label{tableAIC}
\end{table}

\subsection*{Application to neuroscience data}
Application of JS divergence measure (``distance'' between graphs) and its respective statistical test is illustrated in fMRI data of children diagnosed with Attention Deficit Hyperactivity Disorder (ADHD) and children with typical development. ADHD is a developmental disorder that affects at least 5-10\% of children and is associated with difficulty on staying focused, on paying attention, difficulty controlling behavior, and hyperactivity \citep{APA94}. Despite several efforts, there is no comprehensive model of this pathophysiology and the treatment is usually focused on medication that reduces the symptoms and improves functioning \citep{Singh08}.
In order to provide new insights for this disease by using our proposed methodology, pre-processed functional magnetic resonance imaging (fMRI) data, from normal individuals and subjects diagnosed with ADHD, was downloaded from The Neuro Bureau as well as the ADHD-200 consortium (\href{http://neurobureau.projects.nitrc.org/ADHD200/Introduction.html}{http://neurobureau.projects.nitrc.org/ADHD200/Introduction.html}). The data is based on monitoring the BOLD (blood oxygenation level dependent) at different brain regions, which can be considered as an indirect measure of local neuronal activity \citep{Logothetis01}. The data was acquired under a resting state protocol, which is associated with the observation of brain spontaneous activity \citep{Fox05}.

Pairwise Spearman correlation was calculated among 351 mean signals at different regions (using CC400 Atlas, only regions larger than five voxels) and a threshold of p-value = 0.05 (after FDR correction \citep{Benjamini95}) was set to determine the existence of an edge. The correlation between these regions describes the functional connectivity of spontaneous activity at these areas. In other words, an adjacency matrix for each subject was constructed by considering a p-value $<$ 0.05 as 1 and 0 otherwise. Network topological comparisons were carried out between the 478 children with typical development against 158 with combined type of ADHD (hyperactive-impulsive and inattentive).

Differences in the topology between children with typical development and with ADHD were estimated by our approach based on graph spectral distribution and four robust and often used measures, namely number of edges, clustering coefficient, average path length, and degree distribution. The Wilcoxon test was carried out in order to test differences in the number of edges, clustering coefficient, and the average path length. For the degree distribution, we applied the JS based test, similar to the one applied to test differences in the spectra. Table \ref{tableADHD} shows that no statistical evidences to discriminate the two groups of children were identified by the number of edges (p-value = 0.82), clustering coefficient (p-value = 0.85), and average path length (p-value = 0.87). However, by analyzing the degree and spectral distributions (Figure 5), significant statistical differences were found (p-value = 0.031 for degree distribution and p-value = 0.024 for spectral distribution). 

\begin{table}[ht]
\caption{Different metrics to measure graph discrepancy between children with typical development and children with combined type of ADHD (hyperactive-impulsive and inattentive) and their respective p-values. For number of edges, clustering coefficient and average path length, the Wilcoxon test was carried out. For degree and spectral distributions, the JS divergence with the bootstrap test was calculated.}
\centering
\begin{tabular}{c c c c c c}
\hline\hline
& Number of edges & Clustering coefficient & Average path length & Degree distribution & Spectrum\\
normal vs ADHD & 0.82 & 0.85 & 0.87 & 0.031 & 0.024\\
\hline
\hline
\end{tabular}
\label{tableADHD}
\end{table}

\begin{figure}[!tpb]
\includegraphics[angle=0, width=1\textwidth]{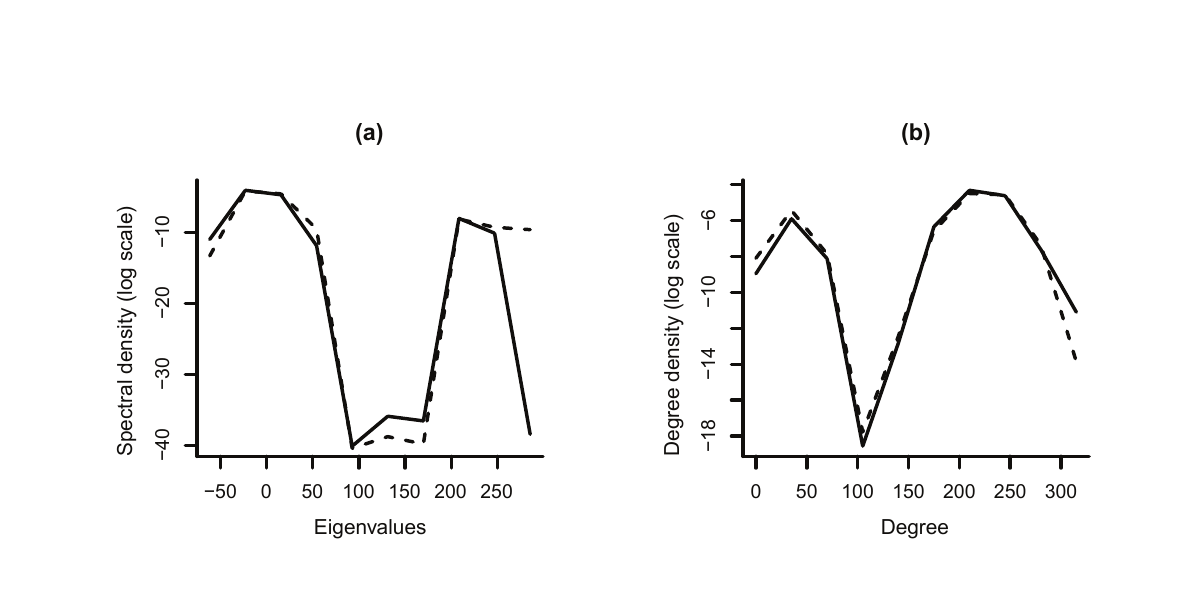}
\caption{(a) Spectral and (b) degree distributions in the log-scale. Solid line represents the children with typical development. Dashed line represents children with combined type of ADHD (hyperactive-impulsive and inattentive).
}\label{fig5}
\end{figure}

In order to check whether the differences in the spectral distributions are not due to numerical fluctuation, the control of the rate of false positives in biological data was verified. The set of 478 children with typical development was split randomly into two subsets, and the JS divergence test in graphs spectra was applied between these subsets. This procedure was repeated 10,000 times. The proportion of falsely rejected hypothesis for p-values equal to 0.1, 1, 5, and 10\% were 0.16, 1.04, 5.55, and 11.05\%, respectively, confirming that the type I error is effectively controlled in this biological data. Moreover, in order to verify the site effect, the JS based test on the spectra was carried out among laboratories. The tests were carried out under the null hypothesis, i.e., in typical development children datasets of different laboratories. Table \ref{tableSiteEffect} shows the p-values after Bonferroni correction for multiple tests. Notice that since no null hypothesis was rejected (significance level of 0.05), there are no statistical evidences of site effect that may significantly affect our results. These results suggest that the differences between children with typical development and with ADHD graphs spectra are statistically significant.

\begin{table}[ht]
\caption{P-values obtained by testing the Jensen-Shannon divergence in the spectra distributions among different laboratories. The tests were carried out under the null hypothesis, i.e., in typical development children datasets of different laboratories. The laboratories were numbered from one to seven and the p-values are after Bonferroni correction for multiple tests.}
\centering
\begin{tabular}{c c c c c c c}
\hline\hline
Labs. & \#2    & \#3    & \#4    & \#5   & \#6   & \#7  \\
\#1   & 0.9   &    1.0   & 0.2   &    1.0    &    1.0    &   1.0   \\
\#2   &          &    1.0   &   1.0   &    1.0    &    1.0    &   1.0   \\
\#3   &          &          &     1.0  &    1.0    &    1.0    &   1.0   \\
\#4   &          &          &           &    1.0    &    1.0    &   1.0   \\
\#5   &          &          &           &          &    1.0    &   1.0   \\
\#6   &          &          &           &         &          &   1.0   \\
\hline
\hline
\end{tabular}
\label{tableSiteEffect}
\end{table}

\section{Discussion} \label{sec:discussion}
The topology of the network represents the set of interactions between the nodes of the network. The topology affects the system's dynamics and carries information about the functional needs of the system, its evolution and the role of each individual unit \citep{Guimera09}. Therefore, network analyses comparing control cases and disease cases is becoming a reference in the medical area \citep{Barabasi11}. Findings of significant differences when doing this comparison will {color{red} possibly} lead to the improvement of diagnostic, prognostic, and therapy. 

Most of the network analyses are based on algorithms that identify punctual changes (presence or absence of a certain edge) in their node connectivity. However, in Systems Biology, different subjects with the same disease may display topologically different molecular networks or brain networks due to genetic variability rather than disease variability. Therefore, a single graph will probably not be representative of the network; instead, a class of graphs generated by a random mechanism seems to be more appropriate.

This situation requires statistical procedures to analyze graphs. The difficulty is then to understand which parameter is representative of the class of graphs. The spectral distribution of a graph gives characteristics for ensemble of graphs generated by the random graphs, and the entropy of a spectrum and Kullback-Leibler divergence between spectra are natural information theoretical quantities to be studied.

	\subsection*{Parameter estimation} 
	
	For some classes of graphs, the parameters of the model can be easily estimated. For example, the parameter $p$ of an Erd\"os-R\'enyi graph can be estimated by counting the edges and dividing it by the total number of possible edges of the graph $(n^{2}-n)$.
However, for more complex models such as the small-world graph proposed by Watts and Strogatz, it is not trivial to estimate the probability $p_{r}$ of edge permutation. Here, we demonstrated that the estimator based on the KL minimum distance (equation \ref{eq:est}) is a general and straightforward method that can be successfully applied to estimate parameters of diverse complex networks.

One may argue whether the application of KL minimum distance estimator could not be applied to degree distribution instead of the graph spectrum. Notice in Figure 3 that the degree distribution showed a lower power to discriminate graphs generated by different parameters than the spectra. Therefore, the spectrum might be a better feature to be analyzed than the degree in order to estimate the parameters.

	\subsection*{Model selection} 

Jeong and others \citep{Jeong00} were the first group to classify protein-protein interaction networks as scale-free graphs by analyzing the degree distribution. Later, several other groups re-analyzed the degree distribution of protein-protein interaction networks and came to differing conclusions regarding whether it was appropriate to refer to these graphs as scale-free \citep{Khanin06, Lima-Mendez09}. One difficulty was the lack of an objective statistical procedure to decide which random graph model fits better the data set.

By applying our model selection approach it is possible to choose objectively, from a choice of candidate graph models, which model best fits the data. By our graph spectrum analysis, all the eight protein-protein interaction networks were classified as scale-free networks among Erd\"os-R\'enyi, scale-free, and small-world models.
 We notice that, in the simulation study, our model selection approach has correctly classified 100\% of the graphs with 120 nodes and the protein-protein interaction networks analyzed here are larger than 700 nodes, which adds to the evidence that among these three candidate networks, the scale-free network seems to fit better.

Despite these results, it is important to notice that the model selection approach is an objective criterion to select the model that best fits the data {\it among candidate models}. Therefore, by analyzing the graph spectrum instead of the degree distribution, this study only provides one more evidence that, scale-free graphs fits better to protein-protein interaction networks than ER and small-world networks. If another complex network model is proposed, one may use this approach to verify which one best fits the given graph.

Another point to be analyzed is the fact that, since only part of the protein-protein network is available, it is always possible that the observed sample is not representative of the entire network, consequently, resulting in a sampling artifact problem \citep{Han05}. Unfortunately, it is a problem about the original data set that should be addressed when the data is collected or by introduction of {\it a priori} model of the network. The analysis proposed here is conditioned to the quality of the data sets.

	\section{Conclusions and future applications} 


Our findings indicate that there are significant differences in the graph spectra of brain networks between children with and without ADHD. We anticipate that future studies in the field of graph spectra may illuminate the topological significance of these features, and consequently help in the investigation of the relationship of these differences with brain function.

The proposed approaches are flexible enough to allow generalizations to other arbitrarily sophisticated families of graphs. Here, we limited the analysis to three well-known classes of random graphs, but the analysis can be extended to other graphs without restriction and it is applicable to many areas where network data is a source of concern.

\section{Materials and Methods}
We present below the details of the computational experiments. The statistical analyses were done using custom made programs in R \citep{R11}(language and environment for statistical computing and graphics).
The R library {\it igraph} was used to generate the random graphs. 

\subsection*{Parameter estimation}
The performance of the parameter estimator based on minimization of the KL divergence was evaluated on different complex network models namely Erd\"os-R\'enyi random graph, scale-free, and small-world, with sizes varying from 50 to 300 nodes. The parameters to be estimated are the probability $p=0.50$, the scaling exponent of the preferential attachment $p_s=1.50$ and the rewiring probability $p_r$=0.30 for Erd\"os-R\'enyi, scale-free, and small-world networks, respectively. The spectral densities ($\rho_{g}$) of each graph were estimated by a Gaussian kernel regression using the Nadaraya-Watson estimator. Since the theoretical spectrum distribution ($\rho_{\theta}$) is unknown for scale-free and small-world networks, the spectrum distribution was estimated by simulating 50 graphs and calculating the average spectra distribution ($\hat{\rho}_{\theta}$) as an approximation for the theoretical distribution ($\rho_{\theta}$). A grid search was carried out in order to determine the argument $\theta$ that minimizes $KL(\hat{\rho}_{g}\vert \hat{\rho}_{\theta})$.

\subsection*{Model selection}
In order to evaluate the performance of the proposed model selection approach, one random graph $G$ is generated (among Erd\"os-R\'enyi, scale-free, and small-world) with parameters $p=0.3$ for Erd\"os-R\'enyi graph, $p_{s}=1$ for scale-free graphs and $p_{r}=0.3$ for small-world graphs, with sizes varying from 10 to 120 nodes. Then, the spectrum of $G$ is estimated. In order to search the optimum set of parameters for each graph model (the set of parameters that minimizes the KL divergence), a grid search was carried out. Fifty graphs for each class ($g_{1}=$ Erd\"os-R\'enyi random; $g_{2}=$ scale-free; and $g_{3}=$ small-world) are generated. The KL divergence is estimated between the spectrum of $G$ and the average spectrum of the 50 graphs of each graph type ($g_{1}$, $g_{2}$, $g_{3}$). The graph model $g_{i}$ ($i=1,2,3$) which has the minimum KL divergence value between $G$ and the three models ($g_{1}$, $g_{2}$, $g_{3}$) is the one which best fits $G$. This experiment was repeated 1,000 times for each graph type (Erd\"os-R\'enyi, scale-free, or small-world) and each graph size (10 to 120 nodes).

\subsection*{Statistical test for JS divergence between graph spectra}
Given two sets of graphs $g_{1}$ and $g_{2}$, the test consists of verifying if the JS divergence between the average graph spectrum of set $g_{1}$ and the average graph spectrum of $g_{2}$ is zero or not. Formally, we test
$H_{0}: JS(\rho_{g_{1}}, \rho_{g_{2}})=0$ versus $H_{1}: JS(\rho_{g_{1}}, \rho_{g_{2}})>0$.

One alternative to perform the test is to use a bootstrap procedure. The {\it bootstrap} was introduced in 1979 as a computer-based method for estimating the standard error of the statistic or to construct confidential intervals that could be used to provide a significance level for a hypothesis test \citep{Efron93}.

Let $\#g_{1}$ and $\#g_{2}$ be the quantity of graphs contained in sets $g_{1}$ and $g_{2}$, respectively. The bootstrap implementation of this test is as follows:
\begin{enumerate}
\item Create a set of graphs spectra $\tilde{g}_{1}$ (the bootstrap sample) by resampling with replacement, $\#g_{1}$ spectra distributions from $g_{1}\cup g_{2}$.
\item Create a set of graphs spectra $\tilde{g}_{2}$ (the bootstrap sample) by resampling with replacement, $\#g_{2}$ spectra distributions from $g_{1}\cup g_{2}$.
\item Let $\rho_{\tilde{g}_{1}^{i}}$ is the $i$-th spectra distribution of $\tilde{g}_{1}$ and $\rho_{\tilde{g}_{2}^{i}}$ is the $i$-th spectra distribution of $\tilde{g}_{2}$. Calculate the average spectra distributions $\rho_{g_{1}^{*}}$, i.e., $\rho_{g_{1}^{*}}(\lambda) = \frac{\sum_{i=1}^{\#g_{1}}{\rho_{\tilde{g}_{1}^{i}}(\lambda)}}{\#g_{1}}$, and $\rho_{g_{2}^{*}}$, i.e. $\rho_{g_{2}^{*}}(\lambda) = \frac{\sum_{i=1}^{\#g_{2}}{\rho_{\tilde{g}_{2}^{i}}(\lambda)}}{\#g_{2}}$,  of $\tilde{g}_{1}$ and $\tilde{g}_{2}$, respectively.
\item Calculate $\hat{JS}(\rho_{g_{1}^{*}}\vert\rho_{g_{2}^{*}})$ (the bootstrap replication).
\item Repeat steps 1 to 5 until obtaining the desired number of bootstrap replications.
\item Test if  $\hat{JS}(\rho_{g_{1}}\vert\rho_{g_{2}})=0$ using the empirical distribution obtained in steps 1 to 5. Gather the information from the empirical distribution of $\hat{JS}(\rho_{g_{1}^{*}}\vert\rho_{g_{2}^{*}})$ to obtain a $p$-value for $\hat{JS}(\rho_{g_{1}}\vert\rho_{g_{2}})=0$, by analyzing the probability of obtaining values equal or greater than $\hat{JS}(\rho_{g_{1}^{*}}\vert\rho_{g_{2}^{*}})$.

\end{enumerate}

The purpose of steps 1 and 2 is to construct new sets $\tilde{g}_{1}$ and $\tilde{g}_{2}$ that are under the null hypothesis. This is exactly done by sampling graphs spectra distributions from $g_{1}\cup g_{2}$. In order to verify whether the bootstrap based statistical test is actually controlling the rate of false positives, p-value histograms under the null hypothesis were constructed. For each class of graph (Erd\"os-R\'enyi random, scale-free, and small-world), 100 graphs with 100 nodes with the same set of parameters ($p=0.5$ for Erd\"os-R\'enyi graphs; $p_{s}=1$ for scale-free graphs and $p_{r}=0.3$ for small-world graphs) were constructed. The 100 graphs of each class were split into two sets of 50 graphs and the statistical test performed with 1,000 bootstrap resampling. These experiments were repeated 10,000 times in order to construct the p-value distributions.

We were concerned in evaluating the power of the proposed test, therefore the parameters of the 50 graphs of one group and the 50 graphs of the other were set with small differences. The parameters are set as follows: $p_{1}=0.50$ versus $p_{2}=0.52$ for Erd\"os-R\'enyi graphs; the scaling exponent $p_{s1}=1.0$ versus $p_{s2}=1.1$ for scale-free networks and $p_{r1}=0.30$ versus $p_{r2}=0.31$ for small-world graphs. The parameters $p_{1}$ and $p_{2}$ for Erd\"os-R\'enyi graphs represent the probability of a pair of nodes be connected by an edge. The parameters $p_{s1}$ and $p_{s2}$ represent the degree of proportionality (scaling exponent) that a new node in the scale-free graph will be connected to node $i$. For example, $p_{s}=1$ means that the new node attaches to node $i$ linearly proportional to the degree of node $i$. $p_{s}=2$ means that the new node attaches to node $i$ quadratic proportional to the degree of node $i$ and so on and so forth. The parameters $p_{r1}$ and $p_{r2}$ represent the probability of rewiring (permuting the edges) in the small-world graph. All other parameters (number of nodes for Erd\"os-R\'enyi graphs and number of nodes and edges for scale-free and small-world graphs) were maintained equal between the two groups.

\section{Acknowledgments}
We would like to thank Adrian M. Bartlett, Christopher Honey, Katlin B. Massirer, Kaname Kojima, and Stephen V. Shepherd for useful discussions during the preparation of this manuscript. We also would like to thank the anonymous referee for carefully reading our manuscript and giving helpful comments that considerably improved the presentation of the article.

\section{Fundings}
DYT was partially supported by Pew Latin American Fellowship and USP project Mathematics, computation, language and the brain. AF was supported by FAPESP grant 11/07762-8 and CNPq grant 306319/2010-1. JRS was partially supported by FAPESP grant 10/01394-4. CEF was partially supported by CNPq grant 302736/2010-7.

\end{document}